\documentclass[aps,prb,reprint,twocolumn,floatfix,superscriptaddress,amsmath,amssymb,amsfonts, nofootinbib]{revtex4-2}

\usepackage{url}
\usepackage{bm}
\usepackage{graphicx}
\usepackage{amsmath}
\usepackage{amstext}
\usepackage{amssymb}
\usepackage{amsfonts}
\usepackage{amsbsy}
\usepackage{verbatim}
\usepackage{color}
\usepackage{braket}
\usepackage[colorlinks=true, urlcolor=blue, linkcolor=blue, citecolor=blue, pdftex]{hyperref}
\usepackage{floatrow}
\usepackage{float}
\usepackage{tikz}
\usepackage{gensymb}
\usepackage{textcomp}
\usepackage{physics}
\usepackage{enumitem}
\usepackage[version=3]{mhchem}
\usepackage{afterpage}
\usepackage{pbox}
\usepackage{dsfont}
\usepackage{siunitx}
\usepackage{stackengine,wasysym,scalerel}
\usepackage{latexsym}
\usepackage{cancel}
\usepackage{comment}
\usepackage{array, multirow, bigdelim, makecell, booktabs}
\usepackage[misc]{ifsym}
\usepackage[capitalise]{cleveref}
\usepackage{times}

\usepackage{color}
\usepackage[colorlinks=true, urlcolor=blue, linkcolor=blue, citecolor=blue, pdftex]{hyperref}

\usepackage[normalem]{ulem}

\definecolor{darkblue}{HTML}{004D6B}
\definecolor{darkred}{HTML}{8c1515}
\definecolor{darkgreen}{HTML}{006400}
\usepackage{hyperref}
\hypersetup{
	colorlinks=true,
	urlcolor=darkblue,
	citecolor=darkblue,
	linkcolor=darkblue,
	breaklinks
}

\newcommand{\dagga}[0]{{\phantom{\dagger}}}

\begin{document}

\title{Emergence of a monopole phase in the $J_1{-}J_2$ Heisenberg model on the triangular lattice for small magnetic fields}

\author{Sasank Budaraju}
\affiliation{Technical University of Munich, TUM School of Natural Sciences, Physics Department, 85748 Garching, Germany}
\affiliation{Munich Center for Quantum Science and Technology (MCQST), Schellingstr. 4, 80799 M{\"u}nchen, Germany}
\author{Shi Feng}
\affiliation{Technical University of Munich, TUM School of Natural Sciences, Physics Department, 85748 Garching, Germany}
\affiliation{Munich Center for Quantum Science and Technology (MCQST), Schellingstr. 4, 80799 M{\"u}nchen, Germany}
\author{Josef Willsher}
\affiliation{Max-Planck-Institut für Physik komplexer Systeme, Nöthnitzer Strasse 38, 01187 Dresden, Germany}
\author{Johannes Knolle}
\affiliation{Technical University of Munich, TUM School of Natural Sciences, Physics Department, 85748 Garching, Germany}
\affiliation{Munich Center for Quantum Science and Technology (MCQST), Schellingstr. 4, 80799 M{\"u}nchen, Germany}
\affiliation{Blackett Laboratory, Imperial College London, London SW7 2AZ, United Kingdom}
\author{Frank Pollmann}
\affiliation{Technical University of Munich, TUM School of Natural Sciences, Physics Department, 85748 Garching, Germany}
\affiliation{Munich Center for Quantum Science and Technology (MCQST), Schellingstr. 4, 80799 M{\"u}nchen, Germany}
\author{Federico Becca}
\affiliation{Dipartimento di Fisica, Universit\`a di Trieste, Strada Costiera 11, I-34151 Trieste, Italy}

\begin{abstract}
We investigate the ground-state phase diagram of the $J_1{-}J_2$ Heisenberg model on the triangular lattice under an external Zeeman field $H$ by using the variational Monte Carlo approach. We span a region with $0 \le J_2/J_1 \le 0.2$ and $0 \le H/J_1 \le 2$, to assess the fate of the (putative) spin-liquid phase that has been detected for $J_2/J_1=1/8$ at zero magnetic field. Simple variational {\it ans\"atze} are proposed for a few candidate states, and their energetics are compared on large clusters to obtain the phase diagram. For $J_2/J_1 \lesssim 1/6$, a continuous transition from a gapless ``Y'' phase to a gapped ``up-up-down'' phase is obtained, as predicted by spin-wave theory. Most importantly, around $J_2/J_1=1/8$, a condensate of monopoles (which are gapless gauge excitations of the spin liquid at $H=0$) is stabilized in a significant region of the phase diagram, for small Zeeman fields. Here, a finite scalar chirality is present, while no transverse magnetic order is detected. The stability of the monopole phase is confirmed by a field-theory approach that includes a self-consistent random-phase approximation of the low-lying spin fluctuations. The boundary between the monopole and ``Y'' phases is also obtained with no free parameters.
\end{abstract}

\maketitle

\section{Introduction}

Triangular lattice antiferromagnets have been a birthplace for key ideas in condensed matter, such as frustrated magnetism and the existence of quantum spin liquids. The latter ones represent exotic states of matter that avoid spontaneous symmetry breaking even at zero temperature and feature emergent gauge fields and fractionalized degrees of freedom~\cite{savary2017,zhou2017,knolle2019field}. It has been suggested that the $J_1{-}J_2$ Heisenberg model hosts a quantum spin liquid around $J_2/J_1=1/8$, although its nature is debated~\cite{zhu2015,hu2015,iqbal2016,hu2019dirac,ferrari2019,jiang2023nature,markus2023,willsher2025a,jiang2026competing,kovalska2026revisiting}. In particular, variational Monte Carlo techniques suggested that the so-called $U(1)$ Dirac state represents a plausible candidate to describe the ground state in the highly-frustrated regime, i.e., $0.08 \lesssim J_2/J_1 \lesssim 0.16$~\cite{iqbal2016}. The low-energy description of this phase is given by massless fermions (spinons) with a cone-like dispersion (i.e., two Dirac points per spin species) interacting with a $U(1)$ gauge field~\cite{ferrari2019,song2019,wietek2024}, which allows for monopole excitations because of its compact nature~\cite{hastings2000dirac,wen2002,hermele2004,hermele2005,hermele2008properties}. In this regard, recent studies have explored the possibility of stable monopoles that do not trigger confinement~\cite{polyakov1977quark}, focusing on their quantum numbers~\cite{song2019,song2020,ganesh2024}. Numerical calculations have also been considered, supporting the possibility that monopoles are gapless in the $U(1)$ Dirac spin liquid~\cite{sasank25mono}.

The external magnetic field represents a standard experimental probe in solid-state physics, often used to probe the susceptibilities of the ground state in the absence of external perturbations. In the case of the $J_1{-}J_2$ Heisenberg model, the external magnetic field is crucial in several aspects. From a purely theoretical side, it allows us to assess the stability of the Dirac spin liquid against the development of long-range magnetic correlations or other kinds of classical order. For example, a transverse $120^\circ$ order may settle, as speculated from theoretical analysis~\cite{ran09ssb} and supported by a recent Schwinger-boson mean-field approach~\cite{dey2024field}. On the practical side, quantitative estimations for the magnetization curve act as an experimental reference to quantify the proximity of a material to the ideal model.

The $J_1{-}J_2$ Heisenberg model with an external magnetic field $H$ has recently received considerable attention. In fact, linear spin-wave calculations~\cite{ye2017quantum, ye2017half} have highlighted the existence of several semi-classical phases, with different spin patterns. More recently, spin-wave calculations have been pushed beyond the linear approximation and density-matrix renormalization group (DMRG) simulations have been implemented on relatively large clusters~\cite{bader2026,keselman2025j_1}; here, a rich phase diagram has been obtained (with some differences between the two works), with evidence that an exotic phase (i.e., not captured by the semi-classical approximation) may exist around $J_2/J_1=1/8$ for small magnetic fields, below the $m=1/3$ magnetization plateau.

Driven by theoretical advances, the search for quantum materials capable of reaching the $J_2/J_1$ window required to have a quantum spin liquid has been carried out~\cite{li2020}. For example, ${\rm AYbSe_2}$ (with ${\rm A=Cs,K,Na}$) and ${\rm YbZn_2GaO_5}$ represent prominent candidates~\cite{scheie2024,xie2023,bag2024}; they host ${\rm Yb^{3+}}$ pseudo-spin-1/2 moments on isotropic triangular lattices where Heisenberg exchange is dominant. Despite their potential, determining the actual value of $J_2/J_1$ is difficult. For example, calculations for ${\rm AYbSe_2}$ using nonlinear spin-wave theory on the $m=1/3$ plateau suggest $J_2/J_1$ ratios between $0.03$ and $0.07$, which may not be enough to reach the spin-liquid regime.
 
In this work, we present a detailed investigation of the phase diagram of the $J_1{-}J_2$ Heisenberg model with an external magnetic field, in the regime $ 0 \leq H/J_1 \leq 2$. In particular, we employ the variational Monte Carlo (VMC) technique based on Gutzwiller-projected fermionic partons~\cite{iqbal2016,sasank25mono}. Our motivation is two-fold: the first one is to obtain an accurate description (using simple variational wave functions) of the conventional ordered states (e.g. Y and canted stripe) that appear in the phase diagram~\cite{starykh2015unusual} and evaluate their region of stability. The second and most important one is to clarify the fate of the Dirac spin liquid when an external field is added to the Heisenberg Hamiltonian. To do this, we extend the construction to include semi-classical phases that appear in the spin-wave approach. For example, the so-called ``Y phase'' (suitable for $J_2/J_1 \lesssim 1/8$), the ``canted stripe phase'' (suitable for $1/8 \lesssim J_2/J_1 \lesssim 1$), and the ``up-up-down phase'' (suitable for the $m=1/3$ plateau). In addition, the ``umbrella phase'' is also considered, as well as a more exotic ``monopole phase'', which is constructed by adding a finite density of magnetic fluxes to the underlying Dirac spin liquid. 

In the zero-field quantum spin liquid, the monopoles lead to critical fluctuations of competing order parameters, and organize the potential instabilities of the gapless spin liquid \cite{hermele2005,hermele2008properties,song2019,seifert2024}. In this work, we examine the possibility that the spin liquid develops a condensate of monopole fluxes under applying an external field $H$, gapping the Dirac cones into emergent spinon Landau levels and confining the gauge degrees of freedom. This effect was proposed by Ran~\textit{ et al.}~\cite{ran09ssb} to lead to a symmetry-broken state with in-plane magnetic order on the kagome lattice, but recent works on the triangular lattice have brought this picture into question \cite{keselman2025j_1,bader2026,wang2026}.

\begin{figure}[t]
\includegraphics[width=\columnwidth]{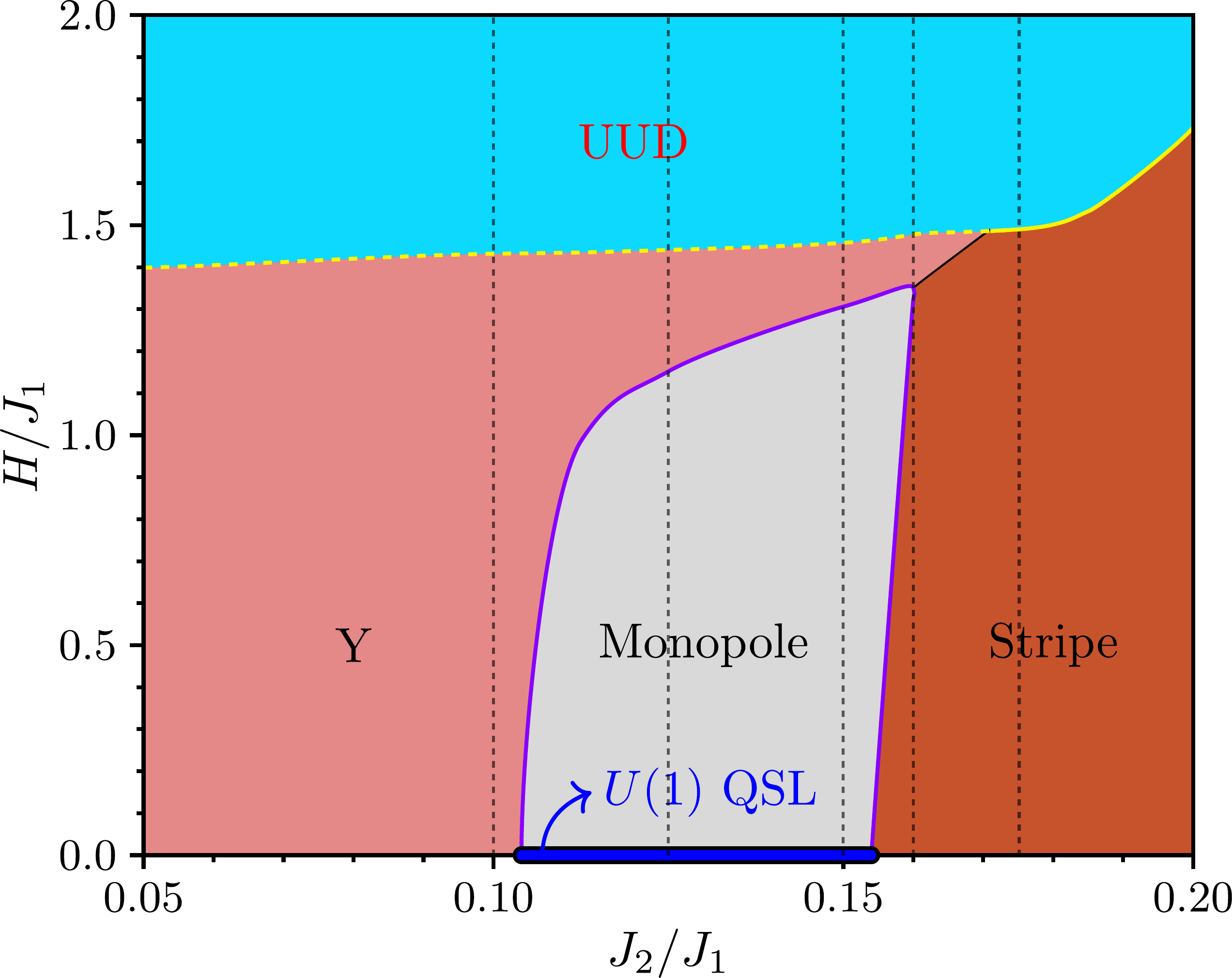}
\caption{\label{fig:phase_diag}
Phase diagram of the model~\eqref{eq:ham} in the parameter range $H/J_1 \in [0,2]$ and $J_2/J_1 \in [0,0.2]$, as estimated from VMC calculations on the $18 \times 18$ triangular lattice. Vertical dashed lines indicate the values of $J_2/J_1$ for which VMC scans were performed (including the case with $J_2=0$). At the phase boundaries, dashed (solid) lines denote continuous (first order) phase transitions. This figure is adapted from Ref.~\cite{bader2026}.}
\end{figure}

The main results of this work are summarized in the phase diagram of Fig.~\ref{fig:phase_diag}. We observe a large portion of the phase diagram where a monopole phase is stabilized over semiclassical ordered states, and first-order transitions to neighboring Y and canted stripe orders. Overall, our results are in good agreement with recent spin-wave theory and DMRG calculations~\cite{keselman2025j_1} on both the extent of the various phases and the nature of phase transitions between them, giving support to our approach. We would like to remark that the present phase diagram is built from calculations on a finite cluster, without a size-scaling analysis; for this reason, the spin-liquid region at $H=0$ is underestimated with respect to the actual values obtained in Ref.~\cite{iqbal2016}. The remarkable outcome is that the monopole phase gives the best variational {\it ansatz} in the highly-frustrated regime and for small external fields and hence represents the natural evolution of the $U(1)$ Dirac spin liquid when the external magnetic field is switched on. 

We bolster this picture with a complementary analytic calculation of the collective spin excitations in the monopole phase. Within this approach, the monopole phase is stable up to a critical field, above which the semiclassical Y order sets in. These results are in good agreement with our numerical calculations.

Finally, we discuss the scaling of the transverse structure factor and in-plane order with magnetic field and system size at different points in the phase diagram. This points to an extended, gapless monopole phase, with an \textit{absence} of transverse magnetic order in the thermodynamic limit. Although further refinements to the variational wave function may yield lower variational energies (with possibly different physical properties), the present results provide a critical benchmark for future studies, also stimulating a reexamination of field-theoretical frameworks.

The rest of our paper is organized as follows: in section~\ref{sec:ansatzes}, we describe the variational wave functions used to represent the various phases in the phase diagram. In section~\ref{sec:results}, we present the VMC results, including the magnetization curves for a few values of $J_2/J_1$ and the properties of the monopole phase. In section~\ref{sec:fieldtheory} we present a field theoretical description of the collective modes in the monopole phase and provide theoretical calculations for the critical field $H_c$ for the transition to the Y phase. Finally, we give some concluding remarks in section~\ref{sec:conclusions}.

\section{Parton construction of competing phases}\label{sec:ansatzes}

We consider the $J_1{-}J_2$ Heisenberg model on the triangular lattice, coupled to an external magnetic field along the $z$ direction:
\begin{equation}\label{eq:ham}
{\cal H} =  J_1 \sum_{\langle i,j \rangle} \mathbf{S}_i \cdot \mathbf{S}_j + J_2 \sum_{\langle \langle i,k \rangle \rangle} \mathbf{S}_i \cdot \mathbf{S}_k -  H \sum_{i} S^z_i,
\end{equation}
where $\mathbf{S}_i=(S^x_i,S^y_i,S^z_i)$ is the spin 1/2 operator on site $i$ and $\langle \dots\rangle$ and $\langle \langle \dots \rangle \rangle$ denote nearest-neighbor and next-nearest-neighbor bonds, respectively. Periodic-boundary conditions are taken on clusters defined by $\mathbf{T}_1=L \mathbf{a}_1$ and $\mathbf{T}_2=L \mathbf{a}_2$ [with $\mathbf{a}_1=(1,0)$ and $\mathbf{a}_2=(1/2,\sqrt{3}/2)$]. Then, the total number of sites is $N=L^2$. Notice that, in presence of a finite magnetic field $H$, the Hamiltonian has only the $U(1)$ (global) spin symmetry, i.e., it is invariant with respect to rotations along the $z$-axis. This leads to the fact that $[{\cal H},S^z]=0$, with $S^z=\sum_{i} S^z_{i}$.

In the following, we explore the phase diagram of the Hamiltonian~\eqref{eq:ham} by constructing variational {\it ans\"atze} for several phases. For that, the first step is to express the spin operator in terms of fermionic degrees of freedom (so-called partons): 
\begin{equation}\label{eq:parton}
S^\alpha_i = \frac{1}{2} \sum_{\tau,\tau^\prime} c^\dagger_{i,\tau} \sigma^\alpha_{\tau,\tau^\prime} c^\dagga_{i,\tau^\prime},
\end{equation}
where $\alpha=x,y,z$ and $(\sigma^x,\sigma^y,\sigma^z)$ are Pauli matrices; the parton operators satisfy the anti-commutation relations $ \{ c_{i,\tau}, c^\dagger_{j,\tau'} \} = \delta_{i,j} \delta_{\tau ,\tau'} $. This rewriting preserves the $SU(2)$ commutation relations of the original spin operators, and is exact if the constraint of one fermion per site is enforced. It also introduces a local $SU(2)$ gauge degree of freedom~\cite{affleck1988}, which enables us to distinguish between different spin liquid states through the projective symmetry group classification~\cite{wen2002,wenbook2004}.

Then, a tight-binding (auxiliary) Hamiltonian for the partons is formulated, with variational parameters optimized to minimize the energy of the physical Hamiltonian~\eqref{eq:ham}. The generic form of this auxiliary Hamiltonian consists of a kinetic hopping term and a site-dependent fictitious magnetic field:
\begin{equation}\label{eq:aux_ham}
{\cal H}_{\text{aux}} = \sum_{\langle i j \rangle, \tau} t_{ij} c^\dagger_{i,\tau} c^\dagga_{j,\tau} + \text{h.c.} + \sum_{i} \mathbf{M}_i \cdot \mathbf{S}_i,
\end{equation}
where hopping parameters $\{ t_{ij} \}$ include specific signs/phases to include the orbital effects of fictitious (local) magnetic fields~\cite{hofstadter1976}; the spin operators must be thought of as expressed by Eq.~\eqref{eq:parton}.

Specifically, we consider four families of variational wave functions, denoted as Y, umbrella, canted stripe, and monopole. The former two have a three-sublattice unit cell, whereas the third one has a two-sublattice unit cell; the monopole states break all lattice symmetries on finite clusters and, therefore, have a unit cell as large as the lattice itself~\cite{sasank25mono}. The details of these {\it ans\"atze} are given below.

\begin{figure}[t]
\includegraphics[width=\columnwidth]{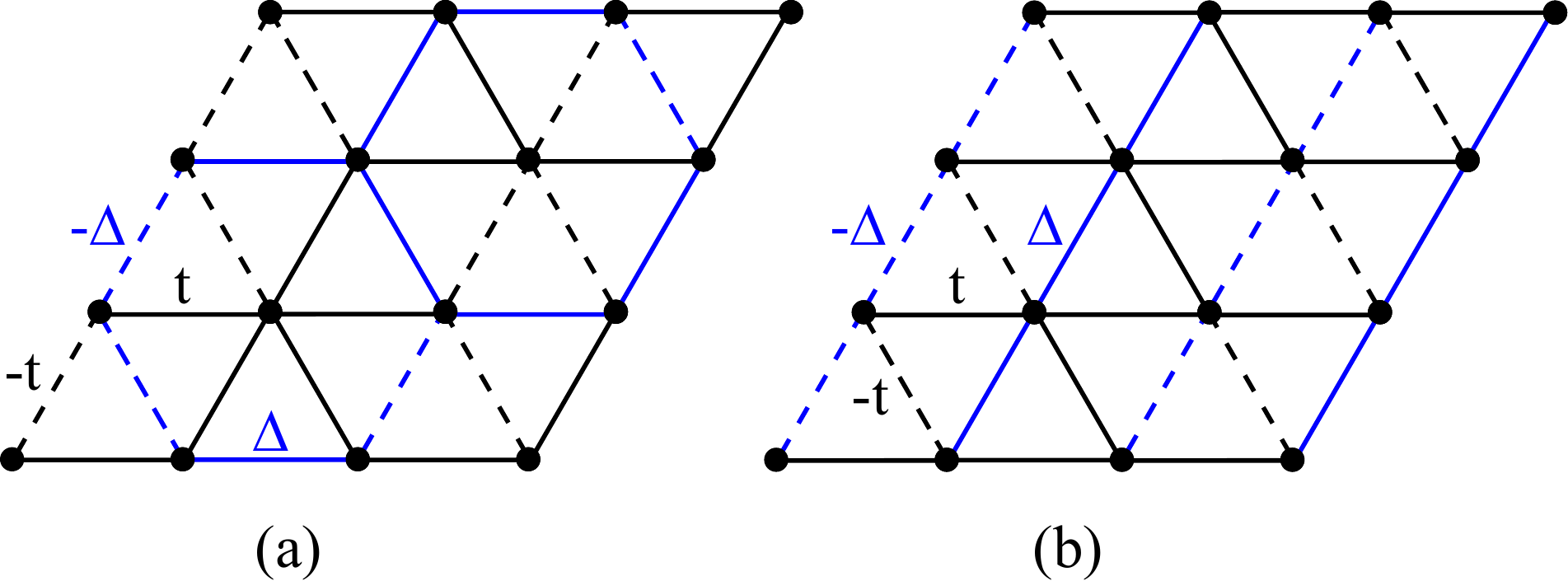}
\caption{\label{fig:ansatzes}
Illustration of hopping parameters $t_{ij}$ for the Y (a) and canted stripe (b) {\it ans\"atze}. Solid (dashed) lines denote positive (negative) hopping amplitudes, and bonds in blue (black) denote amplitudes $|t_{ij}| = \Delta$ ($|t_{ij}| = t$). The flux through each triangular plaquette is $0$ or $\pi$, following the Dirac spin liquid {\it ansatz}~\cite{iqbal2016}.}
\end{figure}

Diagonalizing the auxiliary Hamiltonian ${\cal H}_{\text{aux}}$ yields the single-particle orbitals. Then, the many-body state $\ket{\Phi}$ is defined by filling the $N$ lowest-energy levels. In general, $\ket{\Phi}$ has a non-zero overlap with configurations containing zero or two fermions on one or more sites. The constraint of one fermion per site is achieved by applying the Gutzwiller projector:
\begin{equation}\label{eq:gutz}
{\cal P}_G=\prod_i (n_{i,\uparrow} - n_{i,\downarrow})^2,
\end{equation}
where $n_{i,\tau} = c^\dagger_{i,\tau} c_{i,\tau}$. Furthermore, the presence of a generic field $\mathbf{M}_i$ may mix the spin flavors in the tight-binding model, so that the resulting ground state $\ket{\Phi}$ will not have a well defined quantum number $m$ for the total spin component $S^z$. In our simulations, we choose to work with states that have a well defined magnetization, because this is a conserved quantity of the Hamiltonian~\eqref{eq:ham}. Therefore, we include a projector to a specific sector with
\begin{equation}
m = \frac{2 S^z}{N},
\end{equation}
yielding finally the variational wave function:
\begin{equation}\label{eq:spinwf}
\ket{\Psi} = {\cal P}_m {\cal P}_G \ket{\Phi}.
\end{equation}
We emphasize that the projectors ${\cal P}_m$ and ${\cal P}_G$ are enforced exactly in the Monte Carlo simulation by sampling only the subspace spanned by the allowed configurations (e.g., configurations with one fermion per site, with $S^z=Nm/2$)~\cite{sorella2005wave,beccabook}. 

We now elaborate on the specific form of $t_{ij}$ and $\mathbf{M}_i$ for each of the states mentioned above. They can be thought of as modifications of the Dirac spin liquid {\it ansatz}, which has $\pi$ flux on each rhomboidal plaquette (i.e., $0$ and $\pi$ fluxes on upward and downward triangular plaquettes or vice versa)~\cite{iqbal2016}. Aside from the monopole states, all other states retain this Dirac $[0,\pi]$ flux structure.

\begin{enumerate}
\item Y state: the hoppings $t_{ij}$ are purely real, and their amplitudes are non-uniform, modulated according to Fig.~\ref{fig:ansatzes}(a). The fictitious fields $\mathbf{M}_i$ are chosen to be different for each of the A, B, and C sublattices of the triangular lattice:
\begin{equation} \label{eq:def_Y}
\mathbf{M}_i =
\begin{cases}
(0,0,h_1) & \text{if } i \in A, \\
(h_2,0,-h_3) & \text{if } i \in B, \\
(-h_2,0,-h_3) & \text{if } i \in C. \\
\end{cases}
\end{equation}
The angles among the three spins on each sublattice depend on the values of the three fields $h_1$, $h_2$, and $h_3$ (still, the projections ${\cal P}_m$ and ${\cal P}_G$ may affect the angles obtained at the non-interacting level). In particular, $h_2=0$ yields the collinear up-up-down (UUD) state. We remark that the projector ${\cal P}_m$ ensures that the $U(1)$ symmetry of the Hamiltonian is not broken on any finite lattice, although the Y phase is super-solid~\cite{starykh2015unusual}. The parameters $h_1$, $h_2$, $h_3$, and $\Delta$ are optimized.

\item Umbrella state: the hoppings $t_{ij}$ are purely real, whose amplitudes $|t_{ij}|$ are translationally invariant. The fictitious  magnetic field field is given by
\begin{equation}
\mathbf{M}_i = h \ [\cos(\mathbf{K} \cdot \mathbf{R}_i), \sin(\mathbf{K} \cdot \mathbf{R}_i), 0],
\end{equation}
where $\mathbf{K}=(4\pi/3,0)$ and $h$ is a variational parameter~\cite{iqbal2016}. Performing a projection to a specific $S^z$ sector gives a uniform $z$ component for all spins. The parameter $h$ is optimized.

\item Canted-stripe state: the hoppings $t_{ij}$ are purely real and their amplitudes are modulated according to Fig.~\ref{fig:ansatzes}(b). The fictitious field again has the form:
\begin{equation}
\mathbf{M}_i = h \ [\cos(\mathbf{X} \cdot \mathbf{R}_i), \sin(\mathbf{X} \cdot \mathbf{R}_i), 0]),
\end{equation}
where $\mathbf{X} = (\pi,-\pi/\sqrt3)$ i.e., it is translationally invariant along the $\mathbf{a}_2$ direction, and alternates between $(h,0,0)$ and $(-h,0,0)$ along the $\mathbf{a}_1$ direction. The other two canted-stripe states can also be constructed by taking $\pi/3$-rotations of the present state (these three states are degenerate in energy). The parameter $h$ is optimized.

\item Monopole state: Here, no magnetic fields are present and $t_{ij}$ are complex, with uniform amplitudes and phases that break the translational symmetries of the lattice. The complex phases are chosen so as to have an additional flux of $\pi Q/N$ through every triangular plaquette, where $Q$ is an integer. A specific gauge choice to generate this flux pattern is given in Ref.~\cite{sasank25mono}. The insertion of this flux leads to Landau levels in the fermionic tight-binding spectrum, specifically $2 Q$ levels with zero energy for each spin species~\cite{song2019,sasank25mono}. Then, by filling all the zero modes for one species (say, up), we obtain a unique state with $S^z=Q$. As a consequence, for every magnetization $m$, the variational {\it ansatz} has {\it no free parameters} and, therefore, no optimization is involved. Monopole states are gapless excitations of the Dirac spin liquid, and have been shown to have large overlaps with exact eigenstates of the $J_1{-}J_2$ Heisenberg model on small clusters~\cite{wietek2024}.
\end{enumerate}

\section{Results}\label{sec:results}

The phase diagram of the Hamiltonian~\eqref{eq:ham} is determined by optimizing the {\it ans\"atze} described in the previous section using the stochastic-reconfiguration technique within the standard variational Monte Carlo approach~\cite{sorella2005wave,beccabook} and comparing the energies of the optimized states. We denote the variational energy of a given {\it ansatz} for an external field $H$ as
\begin{equation}
E(S^z, H) = E(S^z,0) - H S^z.
\end{equation}
In fact, since $S^z$ is a good quantum number of the variational state $\ket{\Psi}$, the Zeeman contribution is simply $H S^z$. In our simulations, the parameters in each state are optimized to minimize $E(S^z,0)$ for each $S^z$ sector. Then, the best state for a given $H$ is the one with the optimal ${\bar S^z}$ such that its energy $E({\bar S^z},H) = \text{min}_{S^z}\{ E(S^z,H) \}$. The magnetization curve is then obtained directly as $m(H) = 2 {\bar S^z}/N$.

\begin{figure}[t]
\includegraphics[width=\columnwidth]{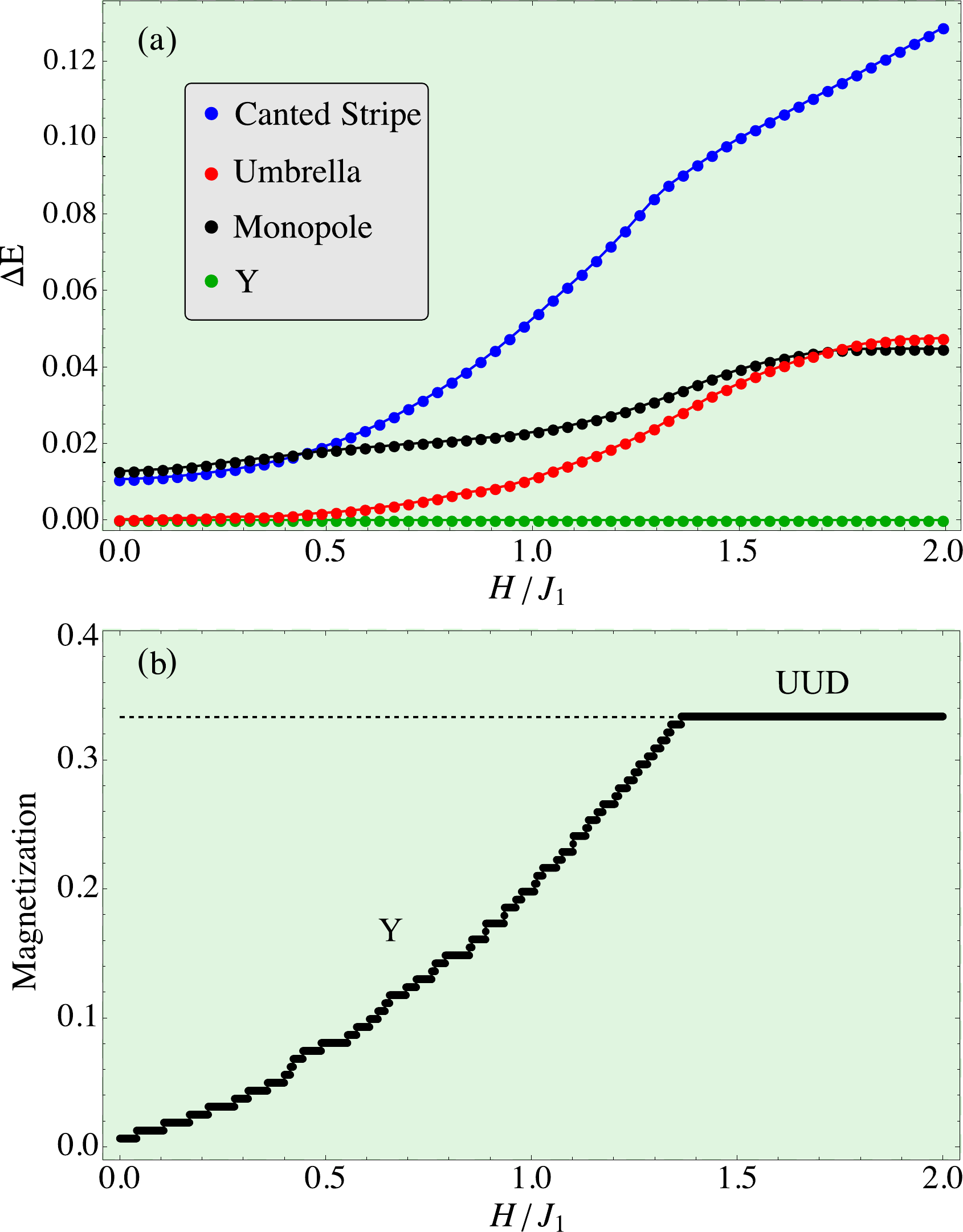}
\caption{\label{fig:J2=0}
(a) Energies (per site) of all the variational {\it ans\"atze} relative to that of the Y phase, on the $18\times 18$ system. Errorbars are smaller than the size of the symbols. (b) Magnetization curve at $J_2=0$, on the same cluster. A continuous transition from the Y phase to the UUD plateau is observed.}
\end{figure}

\begin{figure*}[t]
\includegraphics[width=\textwidth]{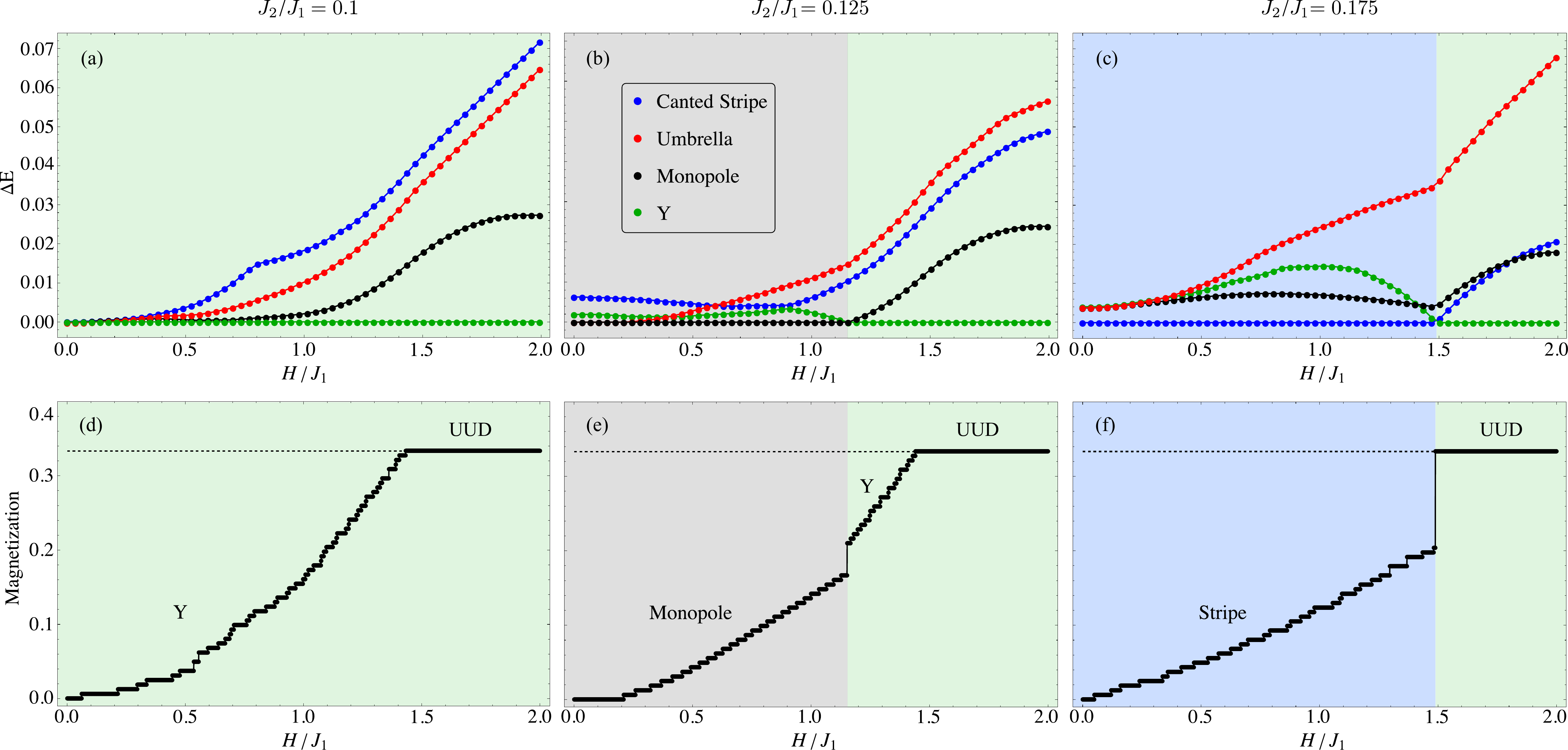}
\caption{\label{fig:otherJ2s}
{ \it Top panels} (a) - (c): Comparison of the energies $\Delta E(H)$ per site for the different {\it ans\"atze} relative to the state with the lowest energy at a given $H$, for $J_2/J_1 = 0.1$, $J_2/J_1 = 0.125$, and $J_2/J_1 = 0.175$. Errorbars are smaller than the size of the symbols. {\it Bottom panels} (d) - (f): Magnetization curves for the three values of $J_2/J_1$. All calculations are performed on a $18\times 18$ lattice.}
\end{figure*}

\subsection{Magnetization and scalar spin chirality}\label{sec:magcurve}

We first discuss the case of the nearest-neighbor Heisenberg model (i.e., $J_2=0$) to validate the method. Indeed, this model has been studied using spin-wave theory~\cite{chubukov1991quantum,ye2017quantum}, and it is well known that the ground state for a small applied field is the Y phase, which, as the field increases, continuously evolves into the collinear UUD state at magnetization $m=1/3$.
 
 The energies of all {\it ans\"atze}, after optimization, are shown in Fig.~\ref{fig:J2=0}(a). The energy of the Y state is lower than that of all other cases throughout the entire region of $H/J_1$ investigated. Although the umbrella state has competitive energies, it is never stabilized in our simulations (for any $J_2$), in agreement with recent non-linear spin-wave theory and DMRG calculations~\cite{bader2026,keselman2025j_1}. The resulting magnetization curve is shown in Fig.~\ref{fig:J2=0}(b). Our numerical results strongly suggest that the transition from the Y to the UUD plateau is continuous, with no jumps of the magnetization up to $m=1/3$, thus confirming previous results~\cite{bader2026,keselman2025j_1}.

Next, we discuss the results for larger $J_2$, namely $J_2/J_1=0.1$, $0.125$, and $0.175$, see Fig.~\ref{fig:otherJ2s}. Remarkably, we find that the monopole states, despite not having tunable parameters, have competitive variational energies throughout the spin-liquid region, lending further support to the underlying Dirac spin liquid in the absence of the external field~\cite{iqbal2016}. At $J_2/J_1=0.1$, the system is close to the onset of the monopole phase: although the Y phase remains the lowest in energy, its energy difference from the monopole states is lower than $0.001$ up to $H/J_1 \approx 0.8$. At $J_2/J_1=0.125$, the monopole states have the lowest energy for magnetizations up to slightly above $m \approx 1/6$, beyond which there is a clear first-order transition to the Y phase. Starting from about $J_2/J_1 \approx 0.16$, the canted-stripe state is stabilized; at $J_2/J_1=0.175$, we observe a direct first order transition to the UUD plateau, indicating the end of the stability of the Y phase. 

Putting all these results together, we have an estimated phase diagram, see Fig.~\ref{fig:phase_diag}. The monopole phase is stabilized for a significant range of applied field and the $J_2$ super-exchange. For the $18 \times 18$ cluster, we estimate that the extent of the phase is $0.1 \leq J_2/J_1 \leq 0.16$. However, we must mention that, as $L$ increases, the monopole gap decreases as $1/L$~\cite{sasank25mono} and, consequently, the stability range of the monopole phase is expected to broaden with increasing $L$.

Our results agree broadly with the predictions of recent spin-wave theory calculations~\cite{keselman2025j_1,bader2026}, both for the extent of the ordered phases and the nature of the phase transitions between them. Our simple variational wave functions thus prove adequate for capturing the essential physics of the model. We also find excellent agreement between our magnetization curves and DMRG data, see appendix \ref{sec:DMRG_comparison}. The VMC {\it ansatz} for the Y phase has only four variational parameters, but achieves an energy comparable to that obtained by DMRG with bond dimension $\chi=50$, whose number of variational parameters scales as $\mathcal{O}(2 N\chi^2) \sim \mathcal{O}(7\times 10^6)$.

\begin{figure}[b]
\includegraphics[width=\columnwidth]{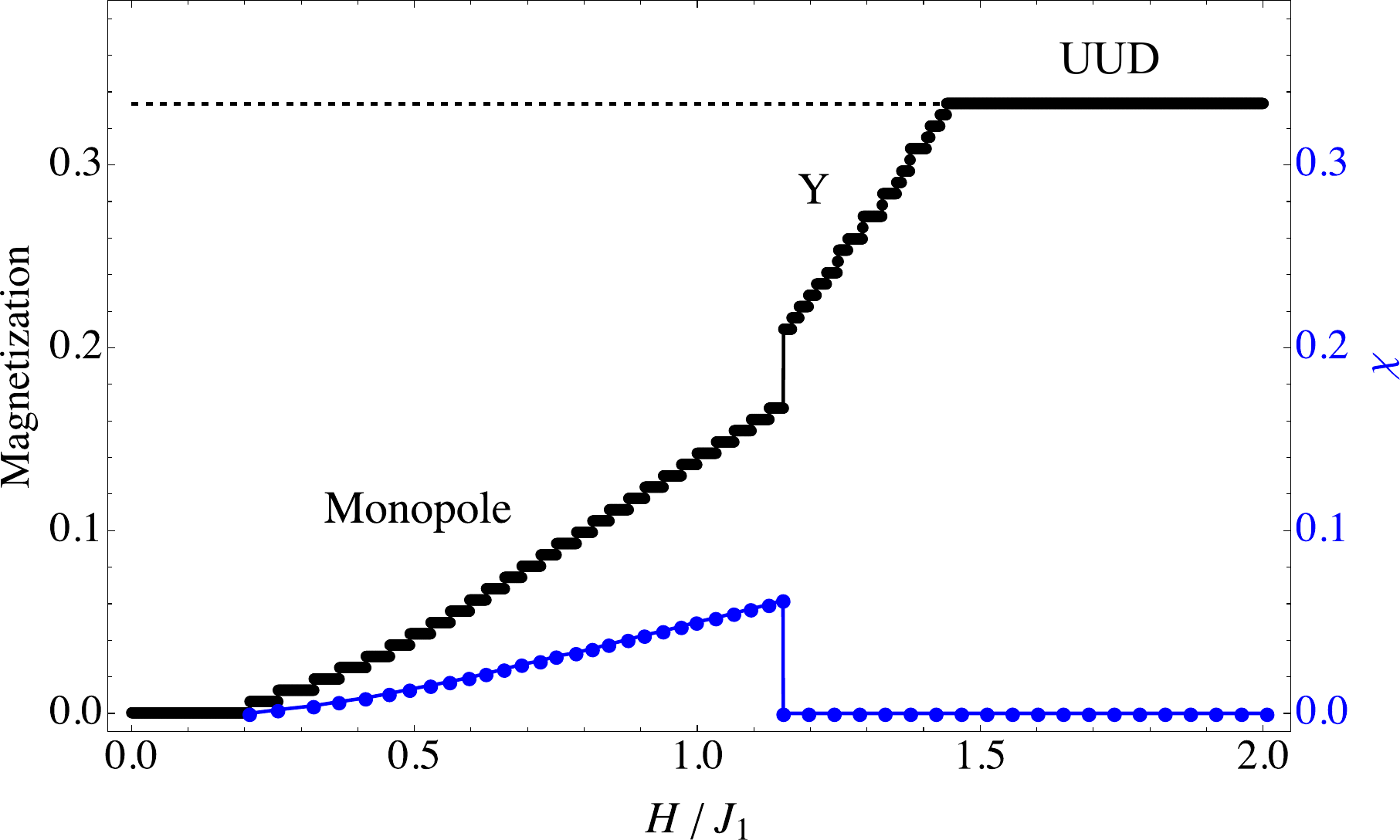}
\caption{\label{fig:scalar_chirality}
Magnetization and scalar chirality at $J_2/J_1=1/8$ as a function of the magnetic field $H/J_1$ on the $18 \times 18$ cluster. The chirality monotonously increases for the monopole phase, and is identically zero for the Y and UUD phases.}
\end{figure}

Finally, the monopole states possess a finite scalar spin chirality (as for the umbrella states). This can be extracted directly from the quantity:
\begin{equation}
\chi = \frac{1}{N} \sum_{[i,j,k]} \mathbf{S}_i \cdot (\mathbf{S}_j \times \mathbf{S}_k),
\end{equation}
where $[i,j,k]$ indicate three nearest-neighbor sites that form an upward triangle in the lattice. The results of the best variational state are reported in Fig.~\ref{fig:scalar_chirality} for $J_2/J_1=1/8$, where the monopole state is stabilized for $H/J_1 \lesssim 1.15$. By contrast, the chirality is identically zero for the Y phase, obtained for $H/J_1 \gtrsim 1.15$, since here the spins are coplanar. This result agrees with the one reported in Ref.~\cite{wang2026}.

\begin{figure*}
\includegraphics[width=\textwidth]{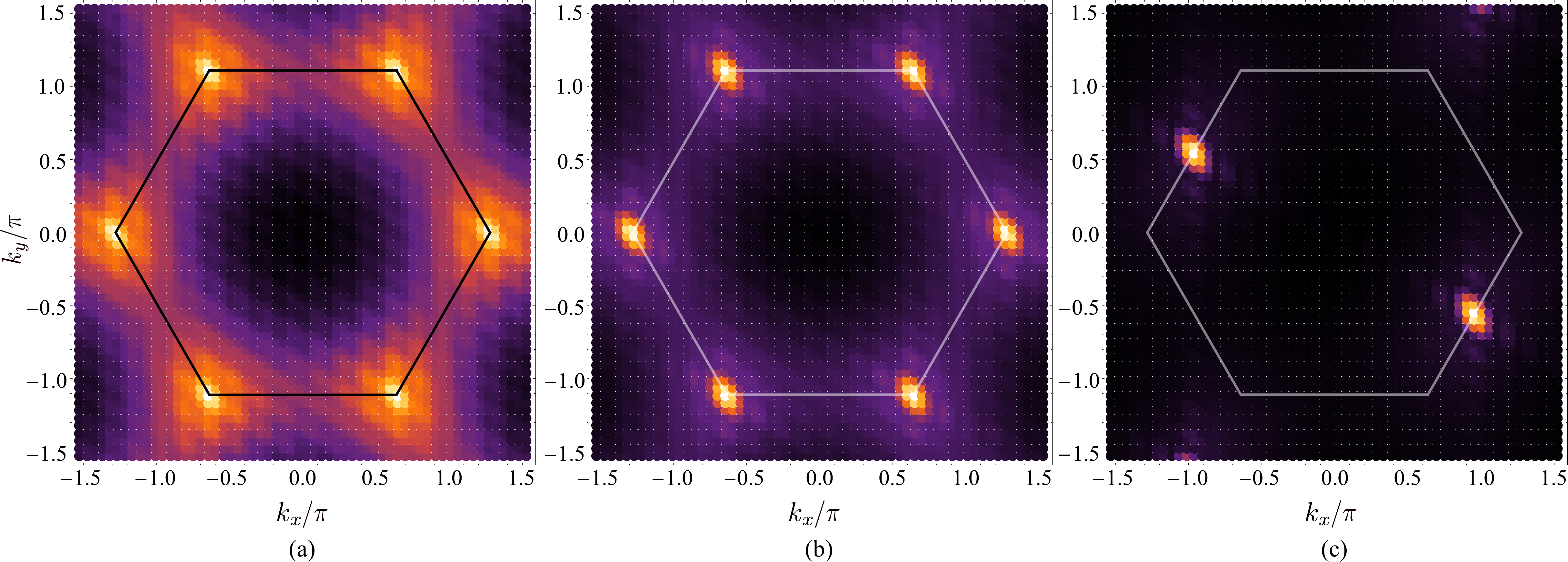}
\caption{\label{fig:S(q)_all}
Static structure factor $S^{xy}(\mathbf{q})$ on the $12 \times 12$ lattice, for (a) a monopole state with finite monopole density, (b) a typical Y state, and (c) a typical canted-stripe state. The results are shown for $m=1/6$.}
\end{figure*}
 
\subsection{Absence of order in the Monopole phase}

We have provided comprehensive numerical evidence that the Dirac spin liquid acquires a finite monopole density in the presence of a magnetic field. This phase was conjectured to have magnetic order in Ref.~\cite{ran09ssb}, as a result of the condensation of the monopoles with lattice momentum  $\mathbf{q}=\mathbf{K}$. Here, we investigate the presence of a possible emergence of magnetic order in the monopole phase by computing the (in-plane) static structure factor:
\begin{equation}\label{eq:sq}
S^{xy}(\mathbf{q}) = \frac{1}{N} \sum_{i,j} e^{i \mathbf{q} \cdot \left( \mathbf{R_i} - \mathbf{R_j} \right)} \langle \left ( S^x_i S^x_j + S^y_i S^y_j \right ) \rangle, 
\end{equation}
where we sum over all pairs of sites $i,j$. A phase with magnetic order at $\mathbf{q}$ is characterized by a structure factor that grows extensively with system size, i.e., $S^{xy}(\mathbf{q}) \propto N$.

\begin{figure}[t]
\includegraphics[width=\textwidth]{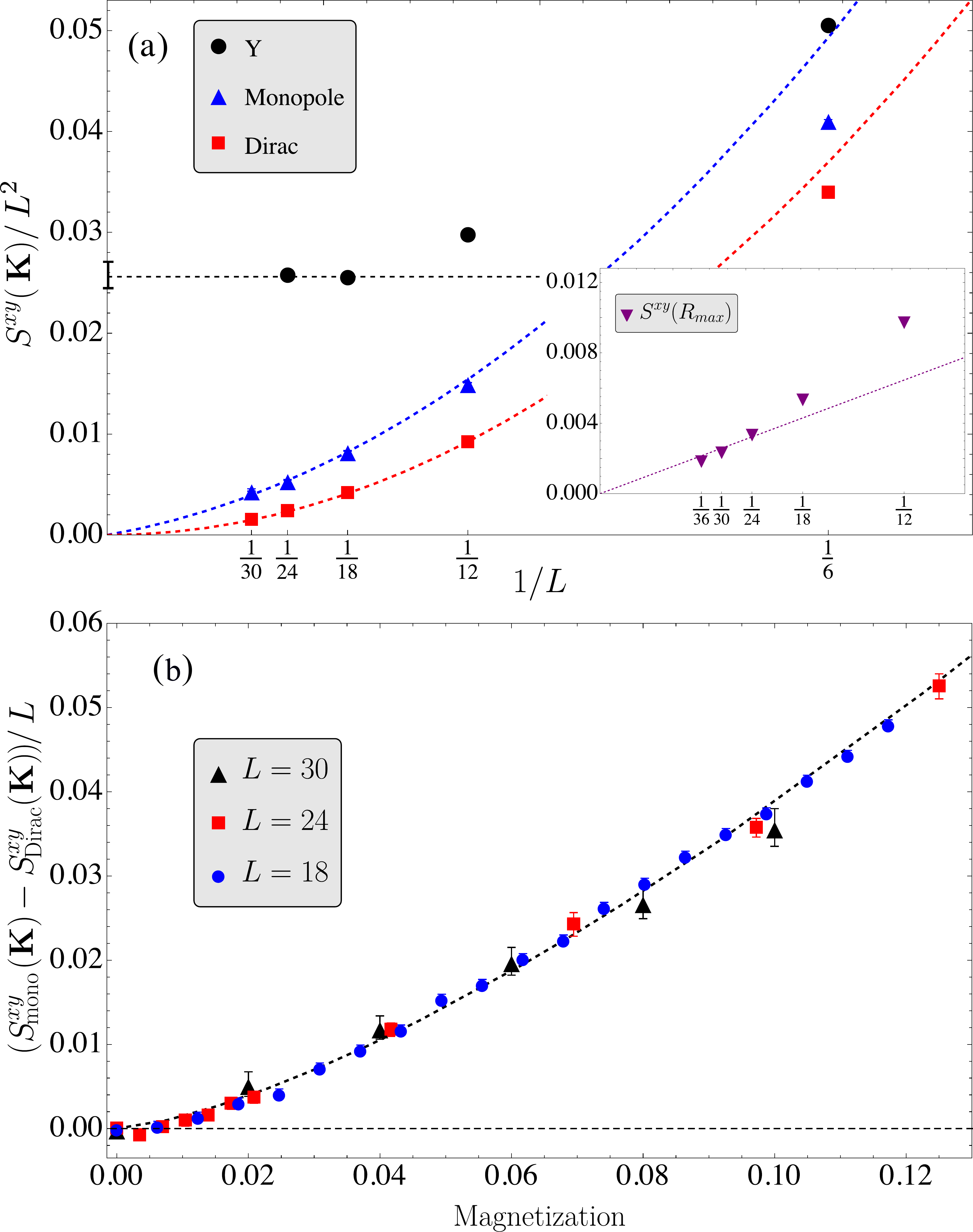}
\caption{\label{fig:SqK_S0Smax}
(a) In-plane structure factor of Eq.~\eqref{eq:sq} at $\mathbf{q}=\mathbf{K}$ divided by $N=L^2$ for the Y and monopole states for $m=1/6$; the Dirac state is also reported for comparison. The red and blue dashed curves represent $1.33/L^2$
and $1.33/L^2 + 0.074/L$, respectively. Inset: spin-spin correlations of Eq.~\eqref{eq:sr} as a function of inverse linear system size $1/L$ for the monopole state for $m=1/6$. (b) Structure factor of monopole states after subtracting the Dirac contribution and renormalizing by $L$, as a function of magnetization for system sizes $L = 18,24,30$. The black dashed line shows the best fit curve,  $1.03 \ m^{1.42}$.}
\end{figure}

In Fig.~\ref{fig:S(q)_all}, we plot $S^{xy}(\mathbf{q})$ for a finite-density monopole state along with a typical Y and canted-stripe state at the same magnetization. The monopole state is chosen to have magnetization $m=1/6$, which corresponds to an additional flux of $\pi/12$ through each triangle. For the Y and canted-stripe states, we observe sharp Bragg peaks at the $\mathbf{K}/\mathbf{K'}$ and $\mathbf{M}$ points, respectively. In contrast, the monopole state features much broader signals around the $\mathbf{K}$ and $\mathbf{K'}$ points, analogous to what is seen in the zero-field Dirac state \cite{iqbal2016}.

In order to understand the ordering characteristics of these different states, we perform a system-size scaling analysis of the Y-ordered state, as well as both the Dirac spin liquid and $m=1/6$ monopole state. We plot the size-scaled in-plane static structure factor $S^{xy}(\mathbf{K})/N$ in Fig.~\ref{fig:SqK_S0Smax}(a). In the case of the Y-ordered state, we confirm that the $S^{xy}(\mathbf{K})/N$ approaches a constant value of $0.026(2)$ in the thermodynamic limit. By contrast, in the case of the (zero-field) Dirac spin liquid, we find that the value of the in-plane structure factor $S^{xy}(\mathbf{K})$ is constant, implying that the scaled quantity $S^{xy}(\mathbf{K})/N$ goes to zero as $1/N$, see Fig.~\ref{fig:SqK_S0Smax}(a).

Turning to the monopole phase, the data appear consistent with the lack of magnetic order on the system sizes considered. The in-plane structure factor $S^{xy}(\mathbf{K})/N$ scales to zero, albeit more slowly than the quadratic behavior of the Dirac spin liquid. Although we cannot exclude the possibility that $S^{xy}(\mathbf{K})/N$ flattens to a small constant value, we highlight that our results are compatible with other works that suggest the absence of in-plane order ~\cite{wang2026}.

To better understand the behavior of $S^{xy}(\mathbf{K})$ in the monopole phase, we study its scaling with magnetization.
For small $m$, the behavior is dominated by an $L$-independent constant, equal to the $m=0$ Dirac state $ S^{xy}_{\text{Dirac}}(\mathbf{K})$. For increasing $m$, we observe an additional contribution, which scales with $m$ and $L$.
We observe the following empirical scaling law:
\begin{equation}\label{eq:smscaling}
S^{xy}_{\text{mono}}(\mathbf{K}) - S^{xy}_{\text{Dirac}}(\mathbf{K}) \propto \ m^{\sigma} L,
\end{equation}
where we estimate the exponent $\sigma = 1.42(3)$. The scaling collapse found by plotting $[S^{xy}_{\text{mono}}(\mathbf{K}) - S^{xy}_{\text{Dirac}}(\mathbf{K})]/L$, as shown in Fig.~\ref{fig:SqK_S0Smax}(b). We interpret the scaling of this quantity in Appendix~\ref{sec:mscaling}, and see this as evidence that the monopole phase is gapless. The in-plane structure factor of the monopole phase $S^{xy}_{\text{mono}}(\mathbf{K})$ has a constant $L$-independent Dirac contribution, which is large on small clusters. The finite density of monopoles leads to a divergent contribution in $S^{xy}_{\text{mono}}(\mathbf{K})$. This contribution diverges slower than $N$, signaling the absence of magnetic order in the monopole phase. This scaling form is extended to the (relatively large-magnetization) $m=1/6$ state and used to produce the blue fit in Fig.~\ref{fig:SqK_S0Smax}(a).

Additionally, the correlations in real space may be considered. To probe magnetic order in the $xy$ plane, we define the spin-spin correlations:
\begin{equation}\label{eq:sr}
S^{xy}(\mathbf{R}_{\rm max}) = \langle S^x_0 S^x_{\rm max} \rangle + \langle S^y_0 S^y_{\rm max} \rangle,
\end{equation}
where ${\rm max}$ corresponds to the site at the maximum distance (with periodic-boundary conditions) to the site at the origin of the cluster, specifically $\mathbf{R}_{\rm max}=L/2 (\mathbf{a}_1+\mathbf{a}_2)$. 
In the inset of Fig.~\ref{fig:SqK_S0Smax}(a), we show that these correlations go to zero in the thermodynamic limit, further suggesting the absence of magnetic order in the monopole state.

\section{Phase transitions to competing orders: Field theory approach}\label{sec:fieldtheory}

In this section, we present a complementary theoretical calculation of the stability of the monopole phase against semiclassical Y order. We find that the monopole phase, which emerges continuously from the zero-field $U(1)$ Dirac spin liquid, is stable for a large region of the $J_2{-}H$ phase diagram.

This theoretical calculation of the phase boundary between quantum spin liquid and semiclassical orders follows the method introduced in Ref.~\cite{willsher2025a}. Here, it was shown that transitions to competing semiclassical orders in the $J_1{-}J_2$ Heisenberg model on the triangular lattice can be determined self-consistently by the condensation of a fluctuating order parameter~\cite{ghaemi2006}, called a paramagnon $\vec{\phi}(x)$. From the spin-liquid perspective, it can be seen as a particle-hole bound state of the fractionalized spinons. In the continuum picture, fluctuations of antiferromagnetic order (with ordering wavevector $\mathbf{K}$) are captured with a continuum field theory of this bosonic field~\cite{affleck1985,affleck1986}.

The zero-field transition from the $U(1)$ Dirac spin liquid to $120^\circ$ coplanar magnetic order is described by the following continuum action:
\begin{equation}\label{eq:hamrs}
S = \int d^3x \left[|\partial_t \vec{\phi}(x)|^2 - c^2|\bm{\nabla}\vec{\phi}(x)|^2 - \alpha^2 r |\vec{\phi}(x)|^2 \right].
\end{equation}
Thus, the bosonic field has a dispersion $\omega=\sqrt{\Delta^2+c^2\mathbf{q}^2}$, with a gap $\Delta^2=\alpha^2 r$ for small momenta $\mathbf{q}$ around the ordering wavevector. We evaluate this gap self-consistently as a function of the Hamiltonian parameters $J_2/J_1$ (i.e., no free parameters). At the critical point $r=0$, the antiferromagnetic order $\vec{\phi}(x)$ condenses and we get a symmetry-broken state. The boson has linear dispersion $\omega=c|{\mathbf{ q}}|$ at the critical point $\Delta=0$ and the transition has dynamical exponent $z=1$. Gauge-field interactions are expected to place it in the QED$_3$ chiral Heisenberg Gross--Neveu universality class~\cite{willsher2025a,dupuis2019}.

Applying a magnetic field $H$ breaks $SU(2)$ spin symmetry, which will split the degeneracy of the paramagnon mode into three $S^z$ eigenstates, with eigenvalues $0,\pm1$. We now study the instability of the magnetized spin liquid state to semiclassical magnetic ordering, caused by the condensation of the paramagnon. Note that we do not assume anything about the nature of the spin-liquid {\it ansatz} in the following, other than the absence of mean-field magnetic order $\mathbf{M}_i$ in Eq.~\eqref{eq:aux_ham} before projection. The lowest-energy state $\phi_+(x)$ with $S^z=+1$ has an effective action~\cite{sachdev2011}:
\begin{multline}
 S = \int d^3 x \left[|(\partial_t-i H)\phi_+(x)|^2 \right . \\ 
 - \left . c^2|\bm{\nabla}\phi_+(x)|^2 - \alpha^2 r |\phi_+(x)|^2 \right],
\label{eq:hamrs_H}
\end{multline}
with dispersion $\omega=\sqrt{\Delta^2+c^2{\bf q}^2}-H\approx (\Delta-H) + c^2 {\bf q}^2 / (2\Delta)$. As such, the semiclassical order condenses at a critical field strength determined by the zero-field paramagnon gap $H_c=|\Delta|$. This assumes that the gap $\Delta$ itself is independent of the field, see Appendix~\ref{sec:appendixfieldtheory} for details. The resulting transition is like a Bose--Einstein condensation of paramagnons~\cite{nikuni2000,giamarchi2008}. The condensed magnetic order has the same unit cell as $120^\circ$ order, but an out-of-plane component $S^z=+1$; hence, it is naturally understood as Y order. Interactions are expected to qualitatively modify the $z=2$ critical point and may drive the transition first order (quartic interactions are marginal and this theory is in the upper critical dimension).

\begin{figure}
\includegraphics[width=\columnwidth]{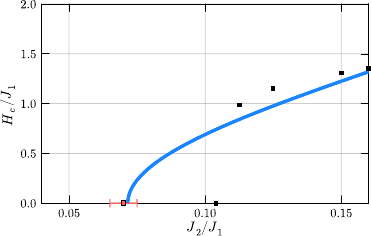}
\caption{\label{fig:theoryfig}
Theoretical prediction of the critical field strength $H_c/J_1$ of the monopole phase to Y-order in the Heisenberg model as a function of coupling $J_2/J_1$. The blue curve is Eq.~\eqref{eq:rpahc}, calculated by measuring the closing of the gap of the paramagnon under an applied field $H$ as a function of coupling $J_2$. The square points mark the transition between the monopole and Y phase as obtained within the VMC method on the $L=18$ cluster. Because of strong finite-size effects in the gapless zero-field limit, we show in red the zero-field transition $0.06<J_2/J_1<0.08$ of Ref.~\cite{iqbal2016}. This is extrapolated in the thermodynamic limit at $H=0$.}
\end{figure}

We are interested in describing the phase boundary to Y-order, which corresponds to a closing of the paramagnon gap at the $\mathbf{K}$ point. Because there are no low energy fermions at this wavevector, we approximate the $\omega=0$ susceptibility of the finite-$H$ monopole state as equal to the Dirac state. The gap $\Delta$ at the momentum $\mathbf{K}$ can be calculated self-consistently in the zero-field Dirac state by integrating out the fermion fluctuations at one-loop level (Appendix~\ref{sec:appendixfieldtheory})~\cite{willsher2025a}.  We then predict the critical field strength $H_c\approx|\Delta|$ as a function of the Hamiltonian parameters $J_2/J_1$,
\begin{equation}\label{eq:rpahc}
H_c/J_1 \approx 2.93\sqrt{\frac{2/3}{1-2J_2/J_1} - 0.778} ,
\end{equation}
where the numerical constants are computed on a large system, e.g., with $L=720$. This calculation is self-consistent, with no free parameters. The critical field strength $H_c/J_1$ is shown in Fig.~\ref{fig:theoryfig}, showing a good agreement with the VMC method for large fields. Instead, for low fields, we find that the self-consistent field-theoretical method predicts a more stable monopole phase than our VMC results on $L=18$ clusters. As discussed in Section~\ref{sec:magcurve}, this can be understood as being due to finite-size effects in the gapless $U(1)$ Dirac spin liquid phase.

Finally, we note that this method cannot be applied so simply to the phase transition to canted stripe order. In this case, the ordering wavevector is at the $\mathbf{M}$ point where the zero-field spin liquid has gapless fermion bilinear excitations. As such, the fermion susceptibility and hence the gap of the paramagnon will depend non-trivially on the applied magnetic field. We leave it to future work to understand how this mechanism leads to the stabilization of the monopole phase over canted stripe order. These results confirm that the monopole condensate phase can be stable to semiclassical ordering for intermediate magnetic fields.

\section{Conclusions}\label{sec:conclusions}

We investigated the triangular Heisenberg model in a magnetic field using variational wave functions and quantum Monte Carlo techniques to optimize the parameters and evaluate the physical quantities~\cite{sorella2005wave,beccabook}. A few relevant {\it ans\"atze} have been analyzed, representing plausible candidate phases for the ground-state phase diagram. In the low-$J_2$ regime, we recover that the Y state is selected below the $m=1/3$ magnetization plateau, where the gapped UUD state is stabilized. For large $J_2$, the canted-stripe state is obtained for small external fields. Most interestingly, in the highly-frustrated regime (i.e., $J_2/J_2 \approx 1/8$, where the same approach has predicted the $U(1)$ Dirac spin liquid at $H=0$~\cite{iqbal2016}), we clearly find the existence of a ``monopole phase'', described by states in which a finite density of magnetic monopoles are added on top of the Dirac spin liquid. 

We provided a self-consistent random-phase approximation (RPA) calculation of the gap to fluctuations of semiclassical Y ordering in this monopole phase, and use it to evaluate a critical field strength that agrees with our VMC calculations. It would be interesting to calculate the instability to (canted) stripe ordering using this method as well, although there exist significant technical challenges in this case, since the semiclassical paramagnon mode sits in the Dirac continuum.

The in-plane structure factor of the monopole phase scales with system size sub-extensively, a fact which points to the absence of magnetic order in the transverse plane (albeit leaves open the possibility of other gapless excitations). This result, which is an incontrovertible property of the variational wave function as constructed here, conflicts with previous theoretical expectations. Of course, we cannot exclude that more refined states, e.g., including linear combinations of monopole configurations at different filling of the zero-energy modes of the auxiliary Hamiltonian, may give a better description of the highly-frustrated regime with $H>0$. However, a numerical treatment that may include these effects is not easily implemented. The absence of magnetic order in the $xy$ plane and the finite scalar chirality represent the hallmark of this state, which is definitely different from classical umbrella states (possessing finite magnetic order in the $xy$ plane). We believe these results will drive further research into the root of the problem, motivating analytical and numerical calculations on frustrated Heisenberg models, as well as experimental investigations on candidate materials.

\section*{Acknowledgements}

We thank Thomas Bader, Urban Seifert and Cristian Batista for helpful discussions. We acknowledge support from the Deutsche Forschungsgemeinschaft (DFG, German Research Foundation) under Germany’s Excellence Strategy (EXC–2111–390814868 and ct.qmat EXC-2147-390858490), and DFG Grants No. KN1254/1-2, KN1254/2-1 TRR 360 – 492547816 [14] and SFB 1143 (project-id 247310070), as well as the Munich Quantum Valley, which is supported by the Bavarian state government with funds from the Hightech Agenda Bayern Plus. J.K. further acknowledges support from the Imperial-TUM flagship partnership.  

\section*{Data availability}
The data presented in this manuscript is available upon reasonable request on Zenodo \cite{zenodo}.

\setlength{\tabcolsep}{12pt}
\renewcommand{\arraystretch}{1.5}
\begin{table*}[t]
\centering
\begin{tabular}{|c|c|c|c|c|c|}
\hline \hline
Magnetization & $J_2/J_1$ & Phase & VMC energy & iDMRG energy & \% Difference \\
\hline 
$1/12$ & $0.08$ & Y & $-0.4938(2)$ & $-0.5059$ & $2.4$ \\
\hline
$1/12$ & $0.125$ & Monopole & $-0.4814(1)$ & $-0.4905$ & $1.8$ \\
\hline
$1/12$ & $0.175$ & Stripe & $-0.4764(2)$ & $-0.4909$ & $2.9$ \\
\hline
$1/6$ & $0.08$ & Y & $-0.4593(2)$ & $-0.4656$ & $1.3$ \\
\hline
$1/6$ & $0.125$ & Monopole & $-0.4426(1)$ & $-0.4530$ & $2.3$ \\
\hline
$1/6$ & $0.175$ & Stripe & $-0.4339(3)$ & $-0.4494$ & $3.4$ \\
\hline
\hline 
\end{tabular}
\caption{Comparison of variational energies obtained by VMC calculations on the $72 \times 6$ torus with iDMRG results on a YC6 cylinder, for a few values of $J_2/J_1$ and magnetization sectors. In the VMC data, the last digit in the parantheses denotes the errorbar. For the DMRG data, the bond dimension is $\chi = 1600$ with the largest truncation error $\sim 10^{-7}$.}
\label{tab:vmc_dmrg_energy}
\end{table*}

\appendix
\section{Benchmarks with the DMRG method}
\label{sec:DMRG_comparison}

\begin{figure}[t]
\includegraphics[width=\columnwidth]{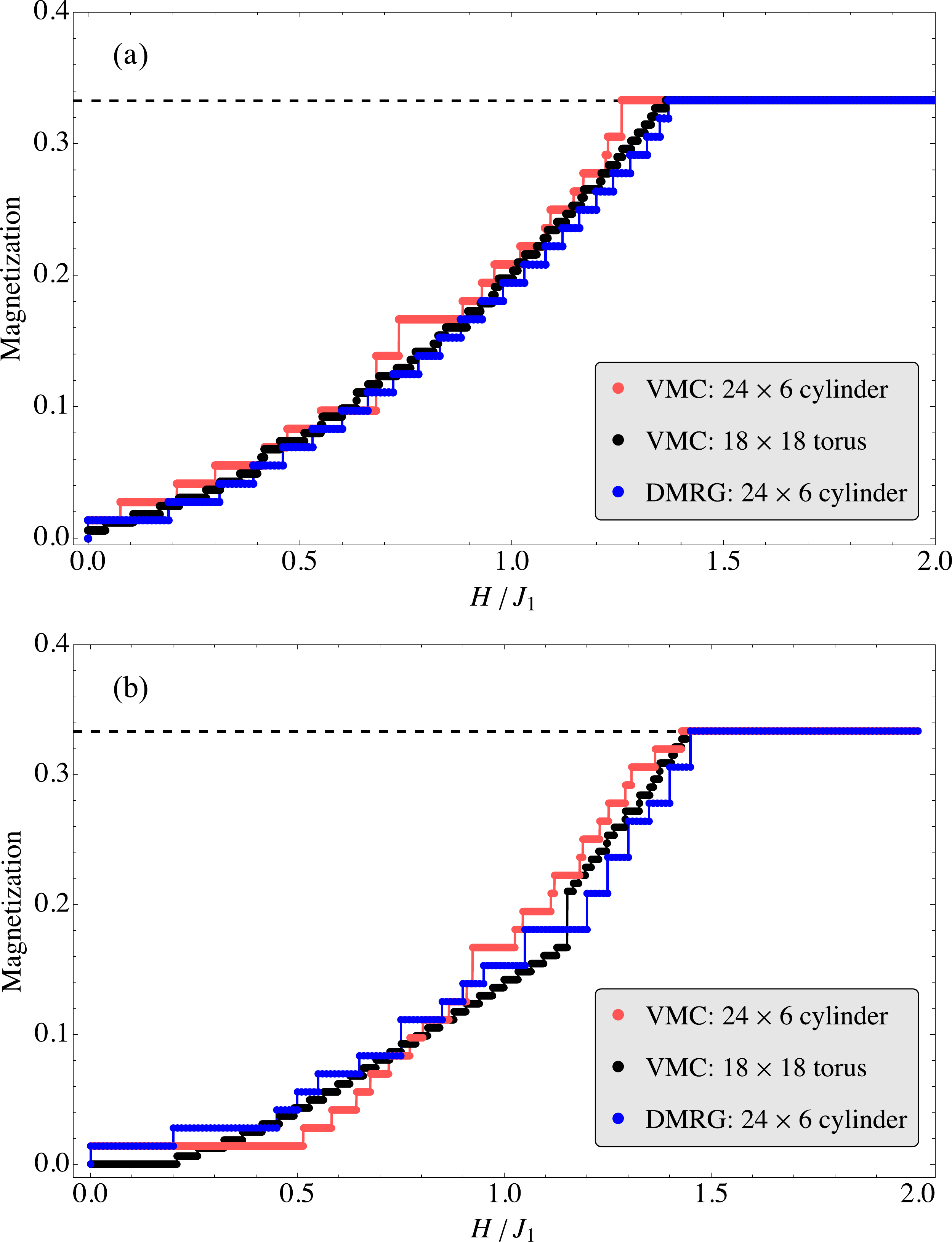}
\caption{\label{fig:vmc_dmrg_j20_mag}
Magnetization curves at (a) $J_2=0$ and (b) $J_2 = 1/8$ obtained by VMC calculations on the $18 \times 18$ torus and a $24\times 6$ cylinder, along with DMRG simulations performed on a $24 \times 6$ cylinder. In the latter case, the bond dimension is $\chi=1600$ with the largest truncation error $\sim 10^{-7}$.}
\end{figure}

Here, we show a few comparisons between VMC and DMRG calculations. The latter approach works best on cylindrical geometries with large aspect ratios, which will be considered here. In Fig.~\ref{fig:vmc_dmrg_j20_mag}(a), we show the magnetization curve at $J_2=0$ obtained by VMC and DMRG simulations. We observe excellent agreement between the different curves, lending strong support to our variational {\it ansatz}, which describes the Y phase. We note that on the $24\times6$ cylinder, the optimized Y {\it ansatz} has a variational energy comparable to that of DMRG with bond dimension $\chi = 50$, which corresponds to around $7\times 10^6$ variational parameters.

In Fig.~\ref{fig:vmc_dmrg_j20_mag}(b), we show a similar comparison at $J_2 = 1/8$. For our VMC calculations on the $24\times6$ cylinder, the monopole phase is never stabilized, and we consequently observe no phase transition. We attribute this to the absence of flux quantization, which is present on a torus. We have verified that a monopole phase is recovered on large asymmetric geometries (e.g., the $72 \times 6$ system), provided that periodic boundary conditions are imposed in both directions. The agreement among the different curves is reasonable, and improves closer to the $1/3$ plateau.

In addition, we also compare the energies of the VMC and DMRG approaches for different values of the frustrating ratio $J_2/J_1$ and magnetization, see Table.~\ref{tab:vmc_dmrg_energy}. In this case, the VMC calculations are performed on the $72 \times 6$ cluster (with periodic-boundary conditions along both directions), while the infinite DMRG (iDMRG) algorithm has been used on the YC6 cylinder~\cite{yan2011}. Typically, we observe that the accuracy of the VMC calculations is always below $3\%$. The best cases are obtained in the putative ``monopole phase'', giving more support for the phase diagram that we obtained. The worst accuracy corresponds to the stripe phase, where our {\it ansatz} breaks the rotational symmetry (on $L \times L$ clusters) and the reflection symmetry on cylindrical geometry. 

\section{Scaling of in-plane structure factor}\label{sec:mscaling}

Here, we take another look at the data shown in Fig.~\ref{fig:SqK_S0Smax} and discuss the scaling of in-plane structure factor with system size $L$ and magnetization $m$. First of all, we notice that the static structure factor is written in terms of the dynamical one as:
\begin{equation}
S^{xy}(\mathbf{q}) = \int_0^\infty d{\omega} \; S^{xy}(\mathbf{q},\omega) \, .
\end{equation}
Whenever the ground state is gapped to some quasiparticle-like excitation at energy $\omega_0>0$, the structure factor takes the form $S(\mathbf{K},\omega)=c\delta(\omega-\omega_0)+S_{\text{inc}}(\mathbf{K},\omega)$, where $S_{\text{inc}}(\mathbf{K},\omega)$ is some incoherent part that vanishes for $\omega<\omega_0$. On finite clusters, $S(\mathbf{K},\omega)$ is finite, with small size effects, such that it converges to a finite value in the thermodynamic limit. Instead, when the system is gapless, the dynamical structure factor possess a continuum, e.g., $S(\mathbf{K},\omega) \approx \omega^{-\alpha}$. On a finite-size cluster, this will be regulated by an infrared cutoff, which scales as $\omega_L \approx 1/L$. In this case, integrating over frequencies can produce a divergent contribution $\omega_L^{1-\alpha} \approx L^{\alpha-1}$ if $\alpha \geq 1$. For $1< \alpha < 3$, this divergence is weaker than $L^2$ and the state is disordered and critical.

\begin{figure}[t]
\includegraphics[width=\textwidth]{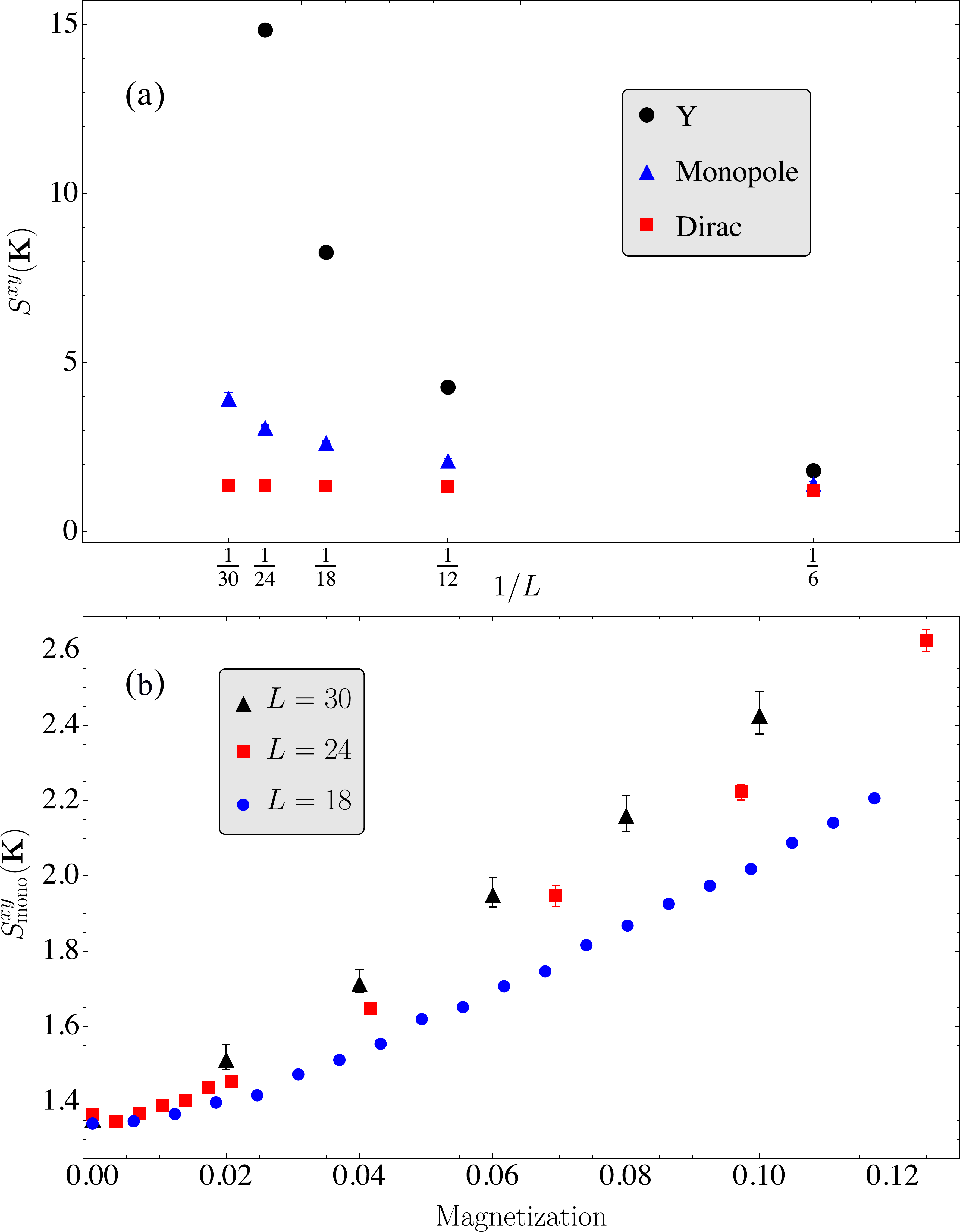}
\caption{\label{fig:sqK_unscaled}
(a) In-plane structure factor of Eq.~\eqref{eq:sq} at $\mathbf{q}=\mathbf{K}$ for the Y, monopole and Dirac states at magnetization $m=1/6$. (b) Structure factor of monopole states, as a function of magnetization for system sizes $L = 18,24,30$.}
\end{figure}

Turning to the $J_1{-}J_2$ Heisenberg model, we present the VMC calculations of the in-plane static structure factor as unscaled data in Fig.~\ref{fig:sqK_unscaled}(a). First of all, the structure factor diverges as $N$ in the Y phase, as discussed in the main text. On the other hand, it goes to a constant in the Dirac state. This is compatible with a finite gap to the paramagnon mode, as discussed in Ref.~\cite{willsher2025a}. We observe that the divergence of the structure factor in the monopole state (with $m=1/6$) behaves as $S^{xy}_{\mathrm{mono}}(\mathbf{K})=S^{xy}_{\mathrm{Dirac}}(\mathbf{K})+cL$, for some constant $c$. This behavior is also seen as a function of magnetization in Fig.~\ref{fig:sqK_unscaled}(b). All together, these observations justify the subtraction of a constant Dirac component in the scaling analysis, performed in the main text in Eq.~\eqref{eq:smscaling}. This result suggests an exponent $\alpha=2$ and a consequently a divergent dynamical structure factor in the monopole phase $S(\mathbf{K},\omega) \sim 1/\omega^2$.

Finally, we note that Ref.~\cite{ran09ssb} predicted this monopole phase has a finite in-plane magnetization $m_{xy}$ due to monopole confinement. Assuming there is a finite in-plane magnetic order parameter, one can predict its scaling as a function of magnetic field $H$ as follows: The magnetization of the mean-field state scales as $m \approx H^2$ and, in the Landau level, the magnetization is proportional to the degeneracy of the Landau level itself, i.e., $m \approx b$, which is linear in the induced flux density $b$. The resulting Landau state has a gapped spectrum with spacing that goes as $\sqrt{b} \approx H$; confinement of gauge fluctuations is then expected to lead to a magnetically ordered state. On the triangular lattice, the $xy$ order parameter is related to the monopole expectation value $m^{xy} \approx \langle \Phi \rangle$ and by dimensional analysis, one can then show that the order scales $m^{xy} \approx H^{\Delta_\Phi}$ where $\Delta_\Phi$ is the scaling dimension of the monopole operator at the QED$_3$ fixed point.

Despite our numerical observations pointing to the absence of magnetic order in the monopole phase, we nonetheless observe a power-law dependence of the divergent contribution on the magnetization
$[S^{xy}_{\text{mono}}(\mathbf{K}) - S^{xy}_{\text{Dirac}}(\mathbf{K})]/L \propto m^\sigma$ [see Eq.~\eqref{eq:smscaling}].
In order to compare with the field-theoretic calculation, we must evaluate this quantity as a function of applied field $H$, but we highlight that this is made difficult by the large $H=0$ plateau.

Preliminary calculations on the $L=18$ cluster point to a scaling $[S^{xy}_{\text{mono}}(\mathbf{K}) - S^{xy}_{\text{Dirac}}(\mathbf{K})]/L\approx H^{2.02}$, which is compatible with a critical exponent $\Delta_\Phi = 1.01(3)$. It is curious that this is compatible with the monopole scaling dimension of QED$_3$, despite there being no in-plane order. We leave it up to future work to evaluate this scaling behavior and to either search for very weak in-plane order, or explain why the field-theoretical treatment breaks down.

\section{Details of the field-theoretical treatment}\label{sec:appendixfieldtheory}

\subsection{Zero-field}

It has been suggested that the finite-energy spectrum of the $U(1)$ Dirac spin liquid is dominated by a sharp spinon-bound state, whose  energy $\Delta$ at the $\mathbf{K}$ point can be evaluated as a function of $J_2/J_1$ in a self-consistent random-phase approximation~\cite{willsher2025a}. Tracking the condensation of this mode as the gap closes ($\Delta \to 0$) allowed for the prediction of a critical transition into the $120^\circ$ magnetically ordered phase.

An effective field theory of the spinon bound state can be written in two steps. First, we form an $O(3)$ field out of the fluctuating order parameters~\cite{affleck1985,affleck1986}. In our triangular lattice model, we focus on an antiferromagnetic order parameter $\vec{\phi}(\mathbf{x}) \sim e^{i \mathbf{K}\cdot \mathbf{x}} \langle \vec{S} \rangle$ with ordering wavevector $\mathbf{K}$. Next, we work in the parton language and, with the approach of Ref.~\cite{willsher2025a}, derive the couplings in the continuum field theory from the lattice model. In practice, we perform a Hubbard--Stratonovich transformation on the interacting spin model in the parton picture to decouple the fluctuations of antiferromagnetic order. Then, we move to momentum space and integrate out the fermions, which gives a self-energy correction to the order parameter field at the one-loop level. Focusing on small momenta $\mathbf{q}$ around $\mathbf{K}$, the non-linear sigma model reads~\cite{willsher2025a}
\begin{equation}
S = \sum_{\mathbf{q}} [- (\alpha^0_{\mathbf{K}})^2(J_{\mathbf{K}}^{-1} - 2\Pi^0_{\mathbf{K}}) +  \omega^2 - c^2{\mathbf{q}}^2 ]| \vec{\phi}_{\mathbf{q}}|^2  \, ,
\end{equation}
up to an overall constant. The mean-field gap is given in terms of the Fourier transform of the Heisenberg interaction
\begin{equation}
J_{\mathbf{K}}^{-1} = \frac{2/3}{J_1-2J_2} \, .
\end{equation}
The constants $2\Pi^0_{\mathbf{K}}=0.778\, J_{1}^{-1}$, $\alpha^0_{\mathbf{K}}=2.93 \, J_{1}^{3/2}$ are evaluated numerically from the mean-field of the $U(1)$ Dirac spin-liquid state. They are defined in terms of the non-interacting fermion susceptibility $\chi^0({\mathbf{q}},\omega)$ as 
\begin{align}
& \Pi^0_{\mathbf{K}} = [\chi^0({\mathbf{q}},\omega)]_{{\mathbf{q}}=\mathbf{K},\omega=0} \, , \\
& \alpha^0_{\mathbf{K}} = 2\left[\frac{\partial^2 \chi^0}{\partial \omega^2}\right]^{-2}_{{\mathbf{q}}=\mathbf{K},\omega=0}\, .
\end{align}
Here, we have related field-theory parameters to the behavior of the fermion susceptibility. Then, we define:
\begin{multline}
\alpha^2 r = (\alpha^0_{\mathbf{K}})^2(J_{\mathbf{K}}^{-1} - 2\Pi^0_{\mathbf{K}}) \\ 
= (2.93)^2 \left[ \frac{2/3}{1-2J_2/J_1} - 0.778 \right] J_1^{\ 2},
\label{eq:alpha2r}
\end{multline}
and we derive the form
\begin{equation}\label{eq:momsp}
S =  \sum_{\mathbf{q}} [-\alpha^2 r + \omega^2 - c^2{\mathbf{q}}^2 ]| \vec{\phi}_{\mathbf{ q}}|^2 \, .
\end{equation}
This leads to a quasiparticle with dispersion
\begin{equation}
\omega = \sqrt{\Delta^2 + c^2\bm{q}^2} \, ,
\end{equation}
with mass $\Delta^2=\alpha^2 r$. The condensation of antiferromagnetic order at $\mathbf{K}$ at $r=0$ is a critical point with dynamical exponent $z=1$. We write this in real-space in the main text Eq.~\eqref{eq:hamrs}.

\subsection{Finite-field}

A uniform magnetic field couples to the ferromagnetic component $m$. This is different to the antiferromagnetic order parameter field; in Haldane's non-linear sigma model description, the uniform magnetization is $\vec{L} = \vec{n}\times \partial_t \vec{n}$. Adding the term $\vec{L}^2 - \vec{H}\cdot \vec{L}$ and integrating out $\vec{L}$ shows that the magnetic field enters as if it were a background gauge field \cite{sachdev2011}. We define the new modified time derivative
\begin{equation}\label{eq:covderphi}
    D_t\vec{\phi} = \partial_t\vec{\phi} - \vec{H}\times\vec{\phi} \, .
\end{equation}
One can write this as a background SU(2) gauge field
\begin{equation}
D_t\phi^a = \partial_t\phi^a  + i H (T^z)_{ac} \phi^{c} \, ,
\end{equation}
where $(T^a)_{bc} = -i \epsilon_{abc}$ are the SU(2) structure constants and the kinetic term looks like a covariant derivative.

We can now rewrite the 3-component field $\vec{\phi}$ in the basis of $S^z$ and raising/lowering operators. This is to simplify the modified derivative of Eq.~\eqref{eq:covderphi}; we define the fields
\begin{equation}
\phi_0 = \phi_z \, , \quad
\phi_\pm = (\phi_x \pm i \phi_y) / \sqrt {2} \, ,
\end{equation}
such that $T^z \phi_\pm = \pm \phi_\pm$.
Then the derivative terms simplify to
\begin{equation}
|\partial_t \phi_0|^2 + 
|(\partial_t - i H) \phi_+|^2 + 
|(\partial_t + i H) \phi_-|^2  \, .
\end{equation}
The mode $\phi_0$ has an effective theory much like Eq.~\eqref{eq:momsp}, but in terms of a single scalar boson.
The action for the other two modes takes the form
\begin{equation}\label{eq:phipmms}
S = \sum_{\bf q} [-\alpha ^2 r + (\omega\mp H)^2 + c^2{\bf q}^2 ]|\phi_\pm|^2 \, .
\end{equation}
These three fields hence have dispersions with the energy given by
\begin{equation}
\omega_0 = \sqrt{\Delta^2 + c^2\bm{q}^2} \, , \quad
\omega_\pm = \sqrt{\Delta^2 + c^2\bm{q}^2} \mp H \, .
\end{equation}
The mode $\phi_+$ with $S^z=+1$ is lowered in energy, and hence will close with a quadratic dispersion. 
To see this, expand the lower band for small momentum ${\bf q}$ away from the minimum at $\bm{K}$
\begin{equation}
\omega_+({\bf q}) \approx (\Delta-H) + \frac{c^2{\bf q}^2}{2\Delta}.
\end{equation}
We can define a chemical potential
\begin{equation}
\mu = \Delta - H = \alpha\sqrt{r} - H
\end{equation}
which controls the transition. When $\mu=0$, the mode condenses with quadratic dispersion; as such the theory now has a dynamical exponent $z=2$. Our RPA calculation Eq.~\eqref{eq:alpha2r} allows us to define a critical field-strength $H_c=|\Delta|$ as a function of $J_2/J_1$. This is discussed in the main text and plotted in Fig.~\ref{fig:theoryfig}. To do this, we make a fundamental assumption that the parameter $\Delta$ is independent of field strength. We expect this to hold for this theory of the transition between the spin liquid state and Y magnetic order, since the fermion spectrum is gapped at the $\mathbf{K}$ point. This means the real part of the fermion susceptibility at zero frequency is well approximated as independent of the field strength. Note that to describe the transition to canted stripe order, this approximation will not hold and we must calculate the field-dependent susceptibility of the Dirac cones. We expect that these effects will act to further \emph{stabilize} the monopole state for the following reason: introducing a gap to the Dirac cone will lower the fermion susceptibility at $\omega=0$. A reduction in $\Pi^0_{\mathbf{M}}$ as a function of $H$ will in turn lead to an increase in $r$, stabilizing the paramagnons. This sketch should be confirmed with quantitative calculations which is beyond the scope of the current work.

\bibliography{refs}
\end{document}